\documentclass[aps,pre,twocolumn,showpacs,amssymb,groupedaddress]{revtex4}
\usepackage{graphicx}

\begin{document}


\title{Lubrication effects on the flow of wet granular materials}
\author{Qing Xu, Ashish V. Orpe, and Arshad Kudrolli}
\affiliation{Department of Physics, Clark University, Worcester,
  Massachusetts 01610} \date{\today}

\begin{abstract}
  We investigate the dynamics of a partially saturated grain-liquid
  mixture with a rotating drum apparatus. The drum is partially filled
  with the mixture and then rotated about its horizontal axis. We
  focus on the continous avalanching regime and measure the impact of
  volume fraction and viscosity of the liquid on the dynamic surface
  angle. The inclination angle of the surface is observed to increase
  sharply to a peak and then decrease as a function of liquid volume
  fraction. The height of the peak is observed to increase with
  rotation rate. For higher liquid volume fractions, the inclination
  angle of the surface can decrease with viscosity before increasing.
  The viscosity where the minima occurs decreases with the rotation
  rate of the drum.  Limited measurements of the flow depth were made,
  and these were observed to show only fractional changes with volume
  fraction and rotation speeds. We show that the qualitative features
  of our observations can be understood by analyzing the effect of
  lubrication forces on the timescale over which particles come in
  contact.

\end{abstract}

\pacs{45.70.Ht, 45.70.Mg}

\maketitle

\section{Introduction} \label{intro}
Granular flows have been studied over the last few decades in order to
confront the challenges posed by several unresolved
questions~\cite{jae96,degen99,raj00}. Most of these studies have
focused on the relatively simple dry granular system with air as the
interstitial fluid. However, lately, there has been a keen interest in
understanding the dynamics of granular flows wetted by a small amount
of liquid considering their prevalence in industry and
nature~\cite{herminghaus05,mitarai06}. The addition of a small amount
of liquid to a dry granular system results in the formation of liquid
bridges which binds the particles.  The static behavior of these
liquid bridges has been studied for quite some
time~\cite{mason65,lian93} on the grain scale and the cohesive force
generated is shown to be dependent on the surface tension of the
liquid and grain size. On the macroscopic scale, there are a few
quantitative descriptions of the impact of the liquid on the overall
behavior.  A wet pile of granular material is more stable than a dry
pile~\cite{hornbaker97,bocquet98} and the failure in a wet pile occurs
in the bulk rather than at the surface as for the dry case
\cite{halsey98,quintanilla01}. Ratholing and jamming can occur during
hopper drainage under conditions where dry grains would normally flow.
Furthermore, particle segregation can be suppressed or enhanced by
making them cohesive~\cite{samadani00,samadani01,li03,li05}. Bond
number (Bo) which is given by the ratio of the strength of the
capillary bridge to the weight of a grain has been introduced as a
useful measure of the cohesivity of the material~\cite{nase99}.

An important system to study granular matter is a partially filled
horizontally rotated drum. This system has been extensively used to
study stability, avalanching, mixing and segregation properties of dry
materials~\cite{rajchenbach90,khakhar97,makse99,ottino00,orpe01,bonamy02,aranson02}.
The flow is mostly confined to the surface and is only a few layers
deep.  A transition from stick-slip to continuous avalanching is
observed as the rotation rate is increased.  More recently, the
rotating drum apparatus is being used to investigate wet granular
materials because it offers many important
features~\cite{quintanilla01,nase99,tegzes02,tegzes03,nowak05}. First,
because the system is periodic, it can be investigated over long times
in a steady state. Second, by simply varying the rotation rate,
various time-scales can be accessed. An attractive practical feature
is that the system can be sealed very easily, which prevents
contamination and evaporation of the liquids.

Recently, using this system, some of the static properties of a wet
pile has been well characterized. Measurements of the angle of maximum
stability as a function of the volume fraction of the liquid, the
grain size, and the dimensions of the wet pile inside the drum have
been made~\cite{nowak05}. Two different models have been developed to
describe the observed dependence. The first phenomenological approach
considers the internal friction of the grains and models the increased
stability due to increase in cohesive force~\cite{mason99,restagno04}.
The second approach considers the geometric stability of the grains
and the increases stability due to the orientation of the capillary
bridges~\cite{nowak05}. Both approaches yield similar scalings for the
maximum stability in terms of the liquid and grain properties which
are observed to be in good agreement with experimental
observations~\cite{nowak05}.

Tegzes {\em et al}~\cite{tegzes02,tegzes03} have performed detailed
measurements of wet granular avalanches in the stick slip regime. They
found that the nature of the avalanches depended on the volume
fraction of the added liquid. The three flow regimes: granular,
correlated, and viscoplastic were characterized by the nature of the
stick-slip avalanches. Recently, simulations have considered the
effect of adding simple cohesive forces on granular
flow~\cite{weber04,brewster05}.

In this paper, we discuss our investigation of the effect of liquids
in the continous avalanche regime. We measure the angle of inclination
as a function of rotation rate of the drum, and the volume fraction
and viscosity of the added liquid. A wide drum is used to minimize the
effect of the side walls. We define a lubrication number which is
given by the ratio of the time needed for the two particles in
neighbouring streamlines to move towards
each other to the time needed to pass each other without contact.
This lubrication
number is found to be useful in describing the observed variations of
the inclination angles.

\section{Experimental Apparatus} \label{sec2}
\begin{figure}
  \includegraphics[width=0.75\linewidth]{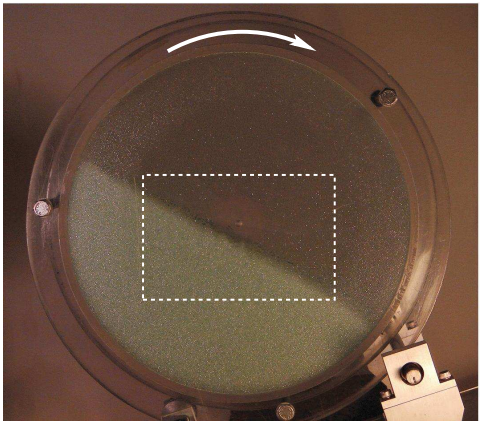}
\caption{An image of the experimental apparatus. The drum is partially 
  filled with a mixture of glass beads and a small amount of silicone
  oil and rotated about its horizontal axis. The dotted box shows the
  region captured in the image to measure the surface angles.}
\label{apparatus} 
\end{figure}

The experimental apparatus consists of a clear plexiglass drum of
diameter $28.5$ cm and width $W = 14.5$ cm rotated about its
horizontal axis as shown in Fig.~\ref{apparatus}.  The drum is rotated
using a computer controlled stepper motor allowing for a rotation rate
($\omega$) variation from $0.00456$ rpm to $13.69$ rpm. The drum is
$40 \%$ filled with spherical glass beads of diameter $d = 1$ mm,
premixed with small amount of viscous liquid.  The amount of liquid
added is reported in terms of volume fraction $V_f$, defined as the
ratio of the volume of liquid to the volume of dry glass beads.  The
volume of dry beads is calculated as the weight of the beads divided
by the bead density ($\rho_{b} = 2.5$ g cm$^{-3}$). The liquid used in
the experiments is Silicone oil with four different viscosities ($\nu
= 0.05, 0.2, 1.0$ and $10.0$ cm$^{2}$s$^{-1}$).
 
Images are taken from one end of the drum and in the central region as
indicated by a dotted box in Fig.~\ref{apparatus}. (The image
resolution of $1024 \times 768$ pixels corresponds to a region of $18$
cm $\times$ 13.5 cm). Images are captured using back lighting to
identify the flowing surface. We have found that such a lighting
scheme is less sensitive to a thin layer of grains which stick to side
walls of the drum and therefore capture the surface of the granular
mixture away from the side walls. The region with grains appear black
in the image and unfilled regions of the drum appear white due to the
back light. The free surface profile is extracted using an edge
detection algorithm from which the angle of the surface is obtained.
Measurements are made for the steady flow achieved after initial few
revolutions of the drum. The reported values for the surface angle are
obtained by averaging over $500$ images (several revolutions of the
drum). During the measurements, the drum is rotating continuously and
particles are always in motion so that the particle-liquid mixture
remains homogeneous throughout. The surface flow of the particles
exhibits a stick-slip motion at lower rotational rates which changes
to continuous motion at higher rotational rates. In the remainder of
the paper, we discuss the flow behavior in the continuous avalanching
regime.

\section{Results and discussion} \label{sec3}
We first discuss the variation in the shape of the flowing surface
with changing rotational speeds and the dependence of surface angles
properties on the choice of location along the flowing surface where
the measurements are made. The effect of interstitial liquid volume
fraction and viscosity on the surface inclination is qualitatively
discussed next. We then examine the relative effect of the liquid
induced forces for various experimental parameters in order to explain
the observed flow behavior.

\subsection{Inclination angle of the flowing surface}
\begin{figure}
  \includegraphics[width=0.75\linewidth]{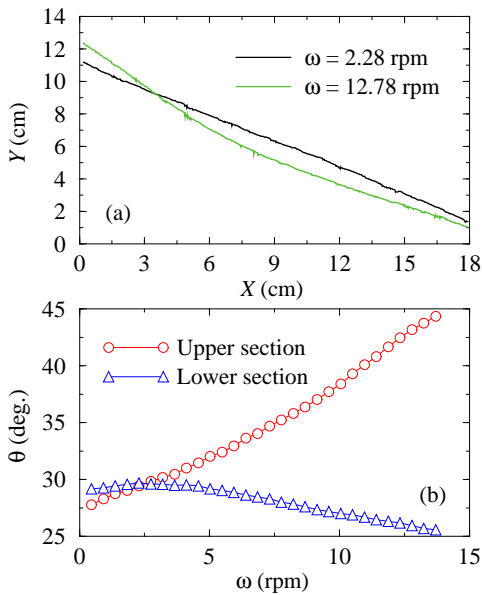}
  \caption{(a) Free surface profile of the flowing particles as
    obtained from the images for two different rotational rates.  (b)
    The surface angles calculated from upper and lower sections of the
    the free surface as a function of rotation rate. The data is
    obtained for the wet system with the added liquid of viscosity
    ($\nu =$ 1.0 cm$^2$s$^{-1}$) and volume fraction ($V_f = 0.48
    \times 10^{-3}$).}
  \label{surface} 
\end{figure}

The surface of the pile is almost linear at low $\omega$ and becomes
progressively non-linear for higher $\omega$. Figure~\ref{surface}(a)
shows a typical surface profile at low and high $\omega$ for wet
systems.  The non-linearity of the surface at higher $\omega$ results
in different slopes along the surface. Figure~\ref{surface}(b) shows
the slopes of the surface in the uppermost and lowermost portion of
the extracted surface profile. The slopes (surface angles) are
obtained by averaging over a region in which the profile is
approximately linear. The surface angle of the upper section
increases with $\omega$ while that in the lower section decreases
slightly with $\omega$. This behavior is observed for dry case as well
and the wet case for all the viscosities and volume fractions of the
liquids investigated. Within the range of experimental parameters
studied, the higher surface angle corresponding to the upper section
appears to be the most sensitive to the variation in system parameters
and in further discussion we focus on this angle to examine the effect
of volume fraction and viscosity of the liquid added to particles.

\subsection{Effect of liquid volume fraction and viscosity} 
\begin{figure}
  \includegraphics[width=0.75\linewidth]{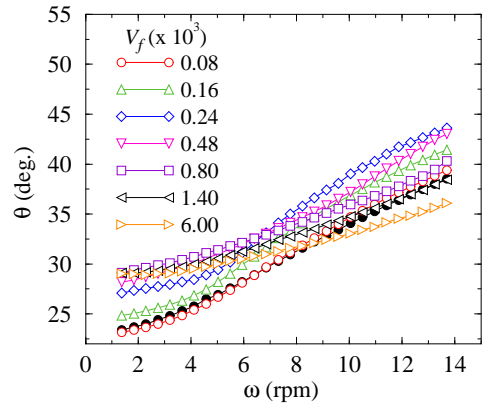}
\caption{(a) Variation of surface angle ($\theta$) with the rotation
  rate for different volume fractions ($V_f$) of the added liquid of
  viscosity $\nu = 1.0$ cm$^{2}$s$^{-1}$. Filled circles denote the
  corresponding data for dry system.}
\label{angle_rot} 
\end{figure}

Figure~\ref{angle_rot} shows the surface angle ($\theta$) plotted
against the rotational speed for different volume fractions ($V_f$) of
the added liquid of viscosity $\nu = 1.0$ cm$^{2}$s$^{-1}$. Also shown
is the data for the dry granular system (filled symbols) for the
entire range of rotational rates studied. For all the cases studied,
the surface angle increases with $\omega$, but the rate of increase
progressively reduces with increasing amounts of liquid added. Notice
particularly the nearly flattened profile at the highest $V_f$.  This
rate of increase with rotational speed for all the volume fractions is
slower than the quadratic dependence observed for the dry systems
\cite{rajchenbach90}. The added liquid tends to bind the particles
strongly resulting in lower surface angles and slower rate of increase
with rotational speeds. A small amount of liquid added ($V_f = 0.08
\times 10^{-3}$) shows a marginal, though a curious change. The angles
for the dry system are higher than the wet system at lower rotational
rate with the trend reversing at higher rotational rates. We attribute
this behavior to the slight fluctuations in the angles for dry system
due to humidity effects which were not controlled in the experiments.
Further increase in the volume fractions results into higher surface
angles than those for the dry case, suggesting the influence of the
acting capillary attractive forces between particles thereby
increasing the resistance to flow. However, adding more liquid also
increases the lubrication between particles causing them to flow much
easily as is seen for the highest volume fraction and higher
rotational rates, where the angles observed are actually lower than
those for the corresponding dry case. Similar behavior is observed for
liquids with different viscosities.

\begin{figure}
  \includegraphics[width=0.75\linewidth]{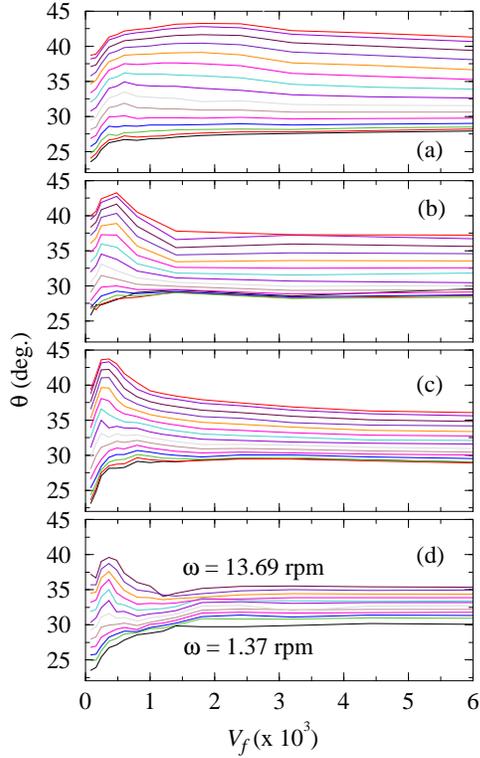}
  \caption{Variation of the surface angle with volume fraction for
    different rotation rates. (a) $\nu = 0.05$ cm$^2$s$^{-1}$, (b) (a)
    $\nu = 0.2$ cm$^2$s$^{-1}$, (c) $\nu = 1.0$ cm$^2$s$^{-1}$ and (d)
    $\nu = 10.0$ cm$^2$s$^{-1}$. In all cases, the rotational rates
    range from $1.37$ rpm to $13.69$ rpm as specified in (d).}
  \label{volfrac} 
\end{figure}

To gain a further understanding about the effect of added liquid, we
replot the data as shown in Figure~\ref{volfrac} for four different
viscosities studied. Consider first the case for $\nu = 1.0$
cm$^2$s$^{-1}$.  For lower rotational speeds, $\theta$ shows a rapid
increase with volume fraction followed by a steady increase over the
remaining range. The increase in the angle with volume fraction is a
direct consequence of the cohesive capillary force due to the liquid
bridges formed between particles as mentioned above. An important
feature which is revealed is the formation of a peak in the surface
angle profiles at higher rotational speeds. The angle rapidly
increases to a maximum and then decreases continuously to a much lower
value at higher volume fractions which is due to the flow being
lubricated by the interstitial liquid. Similar trends are seen for
different viscosities as well. The peak, however, is broad for lower
viscosities and narrower for higher viscosities. A minimum observed at
intermediate volume fractions and the highest rotational rate for $\nu
= 10.0$ cm$^2$s$^{-1}$ is not quite clear and suggests a complex
interplay of the forces due to the liquid.

\begin{figure}
  \includegraphics[width=0.75\linewidth]{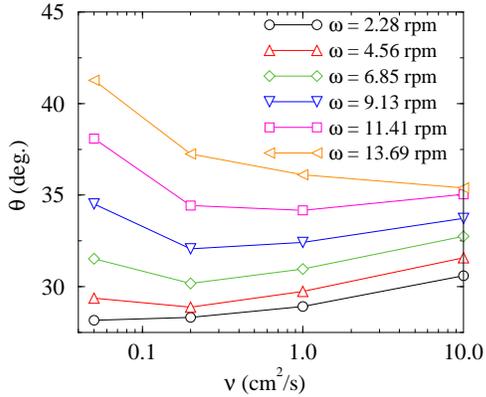}
  \caption{Variation of the surface angle with viscosity for different
    rotational rates and fixed volume fraction ($V_f =$ 6.0 $\times
    10^{-3}$).}
  \label{viscosity} 
\end{figure}

To further elucidate the effect of viscosity, we plot the surface
angles versus the viscosity for different rotational speeds and for
the highest volume fraction studied as shown in
Figure~\ref{viscosity}.  For lower rotational speeds the angles
increase monotonically with viscosity which is expected considering
that the resistance to flow would increase with viscosity. However,
the angles show a reverse behavior at the highest rotational rates,
wherein the liquid actually assists the flow and reduces the angles. A
combination of these effects is seen at intermediate rotational rates
wherein the angles decrease to a minumum and then increase for higher
viscosities.

The discussion so far suggests that the effect of the added
liquid is sensitive to the rotational rates employed.  The primary
effect of adding the liquid is to generate two newer forces between
the particles viz. the attractive capillary forces and the repulsive
viscous forces in addition to the gravity and fricitional forces which
exist in a dry system. In the next section, we examine these forces
and their relative dominance with respect to each other for different
rotational rates which would help in understanding the observed
behavior of the surface angles.

\subsection{Effect of liquid induced forces}
\begin{figure}
  \includegraphics[width=0.75\linewidth]{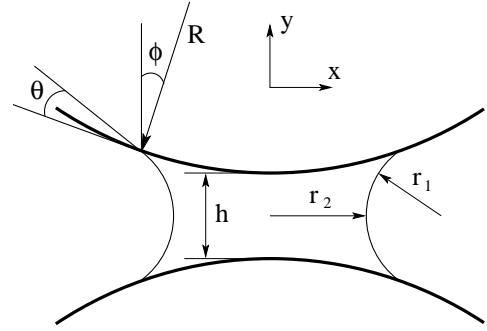}
\caption{Schematic of a liquid bridge formation between two
  spherical particles. $\theta$ is the liquid/solid contact angle,
  $\phi$ is the half-filling angle, $R=$($d/2$) is the diameter of the
  particle and $h$ is the distance between two grains. $r_1$ and $r_2$
  are respectively the radius of the neck and the meridian profile.}
\label{schem-liqbrig} 
\end{figure}

When a small amount of liquid is mixed with dry beads, the liquid wets
the bead system and forms liquid bridges between the particles bonding
them together by a weak attractive capillary force. A schematic of
such a liquid bridge between two particles is shown in
Figure~\ref{schem-liqbrig}. In the limit of small amount of liquid,
the gravity effects can be neglected and the total attractive force
can be expressed as a sum of the component of surface tension
($\sigma$) force along the length of the bridge and the force arising
due to the pressure difference between air and liquid. Further,
assuming $d/2 \gg r_2 \gg r_1$ and $h \ll 2r_1\cos\theta$ and that the
liquid bridge is approximately cylindrical (flat profile), a
simplified expression can be obtained relating the total force to the
bridge volume ($V$) and separation distance ($h$) which is given
as~\cite{pitois00}
\begin{equation}
F_{cap} \approx \pi d \sigma \cos \theta \left [1 - \frac{1}{\sqrt{1
      + \frac{4V}{\pi d h^2}}} \right ].
\label{fcap}               
\end{equation}
where $V = \pi d [ H^{2}(b) - h^{2}] / 4$, $H(r) = h + 2 r^{2}/d$, $r$
is the radial distance from the center of the bridge and $b$ is the
radius of the wetted area. The capillary force varies directly as the
volume of the liquid bridge which depends on the volume fraction of
added liquid.

The pressure change in the liquid bridge arises due to the relative
displacements of two particles with respect to each other while in
flow and the two can be related by the Reynolds equation which can be
simplified to determine the expression for the viscous force acting on
the particles. For finite volumes of liquid (cylindrical bridge) the
viscous force is given as \cite{pitois00}
\begin{equation}
F_{vis} = \frac{3}{8} \pi \rho_{l} \nu d^{2} \left[1 -
  \frac{h}{H(b)}\right]^{2} \frac{1}{h}\frac{dh}{dt}.
\label{fvis}
\end{equation}
where $\rho_{l}$ is the density of the liquid, $dh/dt$ is the rate of
approach of particles towards each other. $H(b)$ is calculated from
the expression given above. The viscous force varies directly as the
rate of displacement of particles, is negligible for large separation
distances and diverges as particles come very close to each other.

\begin{figure}
  \includegraphics[width=0.75\linewidth]{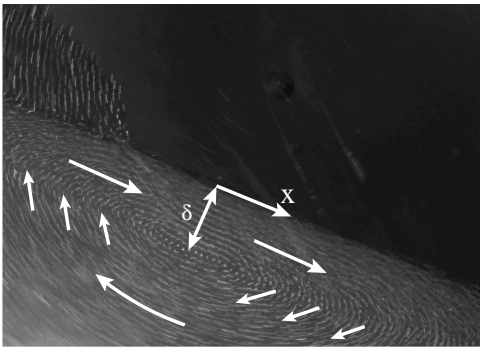}
  \caption{An image taken with 1/4 second exposure shows the depth of
    the surface flow. Volume fraction $V_f=0.08 \times 10^{-3}$,
    rotation rate $\omega$ is 2.28\,rpm. The arrows indicate the
    general direction of the flowing particles.}
  \label{streaks} 
\end{figure}

Direct measurments or estimates of these liquid induced forces in
addition to the gravity and frictional forces for various experimental
parameters would elucidate the observed behavior of surface angles.
However, our experimental techniqe prevents the direct estimates of
these forces between the particles. So instead, we examine different
scenarios determining the relative significance of different forces
with repect to each other. This can be done by calculating the
appropriate timescales corresponding to the motion of particles
relative to each other which can be obtained through velocity profiles
in the flowing layer of particles.  The spreading of the interstitial
liquid on the end plates combined with the sticking of the particles
to the walls prevents accurate measurements of the velocities in the
flowing layer. To get around this, we instead measure the flowing
layer thickness ($\delta$) using streakline photography similar to
that used earlier~\cite{orpe01} for the dry systems. This involves
taking time exposed photographs to obtain particle streaklines as
shown in Fig.~\ref{streaks}. Because we are seeking estimates, we
focus on the central flow region and calculate the maximum layer
thickness over there. We note here that the problems faced while
measuring accurate particle velocities do not hinder the measurements
of the flowing layer thickness. The typical flow direction of
particles and the location for layer thickness ($\delta$) measurement
is shown in Fig.~\ref{streaks}.

\begin{figure}
  \includegraphics[width=0.75\linewidth]{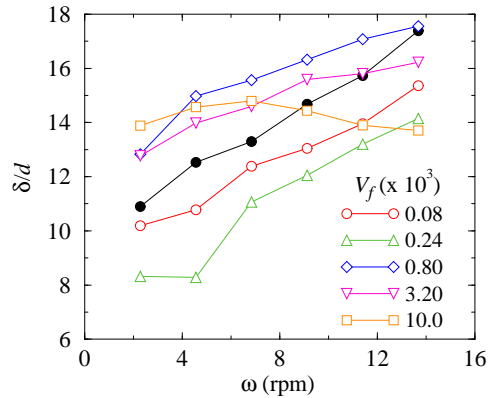}
  \caption{Variation of the normalized flow depth ($\delta/d$)
    with rotation rate for different volume fractions of the added
    liquid ($\nu =$ 0.2 cm$^2/$s), where $d$ is the diameter of the
    particle. Filled circles denote the corresponding data for dry
    system.}
  \label{depth} 
\end{figure}

The variation of the flowing layer thicknes with rotational rate
($\omega$) for different volume fractions and for one particular
viscosity of the liquid is shown in Fig.~\ref{depth}. The layer
thickness changes very slowly with rotational rate, which is similar
to the results for the dry systems with the measurements using NMR
\cite{nak93,yam98} and particles visualization \cite{orpe01}
techniques. The rate of change decreases with increasing rotational
rates, particularly for higher volume fractions with the layer
thickness approaching a constant value at high enough rotational
rates, similar to the trends observed for dry systems. Further, the
layer thickness increases much slowly with rotational rates for higher
volume fractions and interestingly the layer thickness at the highest
volume fraction becomes almost constant and independent of the
rotatational speed. Similar behavior is also observed for liquids with
different viscosities. For the steady flow of particles in the
continuous flow regime, the flux $Q$ of the particles (at the central
location where $\delta$ is measured) can be written from mass
conservation as
\begin{equation}
Q = \frac{1}{2} \rho_b (\omega z R_0) R_0 = \rho_b V_x z \delta.
\label{massbal}
\end{equation}
where $\rho_b$ is the density of glass beads (assumed to be
approximately equal in the flowing layer and rotating bed), $R_0$ is
the radius of the drum, $z$ is the length of the cylinder and $V_x$ is
the average velocity in the flowing layer. The average velocity in the
flowing layer can be then written as
\begin{equation}
V_x = \frac{\omega R_0 ^2}{2 \delta}.
\label{avgvel}
\end{equation}
Using Eq.~\ref{avgvel}, we can thus get the estimates of the
representative velocities in the flowing layer for various rotation
rates, volume fractions and viscosities of the interstitial liquid.
For the particles flowing in the $x$-direction (see
Figure~\ref{streaks}), we denote $t_x = d/(V_x)$ as the time required
for the two particles in the neighbouring streamlines to travel past
each other, where $d$ is the particle diamete r. Substituting for
$V_x$ from Eq.~\ref{avgvel} gives
\begin{equation}
t_x = \frac{2 d \delta}{\omega R_{0}^2}.
\end{equation}
The timescale $t_x$ primarily varies inversely as the rotational rate
($\omega$) since $\delta$ is only weakly dependent on $\omega$ and
volume fractions as seen from Figure~\ref{depth}.

We now define $t_y$ as the timescale for the same two particles to
approach each other while squeezing out the liquid from between.
Balancing the viscous force between the two particles given by
Eq.~\ref{fvis} to the weight of the particle ($\pi d^3 \rho_b g/6$)
and then integrating the resulting equation gives
\begin{eqnarray}
t_{y} &=& \frac{9}{4} \frac{\nu \rho_{l}}{d g \rho_b}
\left[\frac{2 b^2 d (h_s - h_0)}{(2 b^2 + h_s d)(2 b^2 + h_0 d)}
\right.  
  \nonumber \\
&& + \left. \ln\left(\frac{1 + 2 b^2/h_s d}{1 + 2 b^2/h_0 d}\right)
 \right]  
\end{eqnarray}
where $h_s$ is the surface roughness of the beads, and for the glass
beads is approximately $10^{-3}d$. $b$ is the radius of the wetted
area of the particle. The above equation gives an estimate of the time
required by the two particles separated by a distance $h_0$ to
approach each other to a distance within surface roughness ($h_s$).

The ratio $Ra = t_x/t_y$, after substituting for the experimental
parameters, can be then written as:
\begin{equation}
Ra = \frac{0.13\;\; \delta/(\omega \nu)}
{\left[\frac{2 b^2 d (h_s - b)}{(2 b^2 + h_s d)(2 b^2 + b d)} +
  \ln\left(\frac{1 + 2 b^2/h_s d}{1 + 2 b/d}\right)\right]}
\label{Ra}
\end{equation}
where, we have assumed that $h_0 \sim b$, i.e. liquid bridges just
before or after snapoff are semi-spherical droplets on the surface of
the beads. We can now investigate various scenarios under which
different forces can be dominant. For $Ra \ll 1$, the two particles
pass each other without coming in contact (thus minimizing the
frictional forces) while for $Ra \gg 1$ the viscous forces between the
particles dominate the overall flow behavior. In the following we now
estimate these time scales for different rotational rates, volume
fractions and viscosities of the liquids used in our experiments.

For small $V_f$, as when the beads first get uniformly coated with the
liquid, and liquid bridges are formed between the grains, $b \sim
h_s$~\cite{mason99}. In that limit, both the terms in the denominator
of Eq.~(\ref{Ra}) tend to 0 and therefore $Ra \gg 1$ and particles
touch as they flow past each other and kinetic friction/viscous forces
become important resulting in the higher surface angles. This flow
regime corresponds to the surface angles increase with volume fraction
(see Fig.~\ref{volfrac}) leading to a maximum. The peak in the profile
would in that case correspond to the maximum value of $Ra$.

On the other hand, for large enough volume fractions, $b \sim d/2$,
$h_0 \sim d/2$ and using $h_s \sim 10^{-3}d$ and $\delta \sim 14d $
(from Fig.~\ref{depth}), we get $Ra \rightarrow 10^{-2} /(\omega
\nu)$.  For highest rotational rate ($\omega = 13.69$ rpm) and lowest
viscosity ($\nu = 0.05$ cm$^2$ s$^-1$), $Ra \rightarrow O(10^{-1}) <
1$. We thus expect the surface angles to be lower and is indeed seen
from the decrease in surface angles with volume fraction (see
Fig.~\ref{volfrac}(a)). This effect is more pronounced at the highest
viscosity used ($\nu = 10.0$ cm$^{2}$ s$^{-1}$) for which $Ra
\rightarrow O(10^{-3}) \ll 1$.  Note the faster decay in the surface
angles with increasing volume fractions in Fig.~\ref{volfrac}(c,d).
The rotation rate and viscosity, thus, act together to give maximal
lubrication. This is reflected in the non-intuitive decrease in the
surface angles with increasing viscosity at highest rotational rate
(see the topmost curve in Fig.~\ref{viscosity}).

The simple analysis presented here based on the time scales for the
particles to come in contact serves to explain most of the effects due
to varying volume fractions of the liquid for different rotational
rates and viscosities. The primary dependence on the viscosity of the
liquid or the rotational rates for a given volume fraction, however,
cannot be explained using this analysis.  For example, the angle
increases with viscosity at lower rotational rates, the reason being
the increase in the drag forces with viscosity leading to additional
resistance to flow and hence higher angles (see the bottom curve in
Fig.~\ref{viscosity}). Using Eq.~(\ref{Ra}) alone in this instance
would cause the corresponding $Ra$ values to change from $O(1)$ to
$O(10^{-2})$ suggesting a decrease in the angle contrary to that
observed. A similar argument also holds true for the case where the
angles increase with rotational rates (the corresponding $Ra$ actually
decreases in those instances). Apart from these behaviors, the
analysis cannot as well account for the more complex behaviors like
the sharp minimum in the surface angles observed at higher rotational
rates and highest viscosity (see Fig.~\ref{volfrac}d) or the decrease
in the angle followed by an increase with increasing viscosity
observed for intermediate rotational rates (see Fig.~\ref{viscosity}).
Understanding these complicated effects would require a more rigorous
study involving estimation of the different competing forces which is
beyond the scope of this work.
 
\section{Summary} \label{sec6}

In summary, we have investigated the effect of volume fraction,
viscosity of the liquid, and rotation rate of the drum on the
inclination angle of a glass bead pile in a rotating drum. The surface
angle show a non-linear increase with the rotational rate for any
particular liquid volume fraction and viscosity. For any rotation rate
and liquid viscosity, adding small amounts of liquid increases the
surface angle than the corresponding values for the dry system which
is a direct consequence of the liquid bridge formation between
particles thus increasing the resistance to flow. However, continuing
the addition of liquid eventually leads to a constant slow increase in
the surface angles at low rotational rates whereas the angles show a
surprising decrease at high enough rotation rates even though the
strength of liquid bridge does not decrease. The effect is seen to be
more dramatic for higher viscosities. By doing an analysis of the time
scales related to the approach of two particles towards each other
during flow using we derived a criterion for determining the relative
effects of friction and lubrication between flowing particles. The
analysis is valid primarily for accounting the effects due to varying
liquid volume fractions and captures most of the observed variation in
the angle of inclination. However, we are unable to make simple
comparisions of the scaling because of different competing forces in
the experiments which are not measurable using the techniquies
employed here.  Additionally, it would be helpful to have detailed
velocity profiles and flow depths away from the sidewalls as a
function of liquid content, and will be the focus of future work.

\section{acknowledgments}
We thank J. Norton for his help with building the apparatus, and J.
Xia and H. Wang for helpful discussions. This work was supported by
the National Science Foundation grant number CTS-0334587 and the
donors of the Petroleum Research Fund.

\end{document}